\documentclass[%
 reprint,
 amsmath,amssymb,
 aps,
]{revtex4-2}
\usepackage{graphicx}
\usepackage{dcolumn}
\usepackage{bm}
\usepackage{chemformula} 

\usepackage[T1]{fontenc}
\usepackage{amsmath}
\usepackage{newunicodechar}
\usepackage[utf8]{inputenc}
\DeclareUnicodeCharacter{2212}{\textminus}
\DeclareUnicodeCharacter{0308}{\"{}}
\usepackage{xcolor}
\usepackage{hyperref}
\usepackage{siunitx}

\usepackage{xcolor}
\begin{document}

\preprint{APS/123-QED}

\title{Nontrivial Boundary-Mediated Superconducting Transport in a TRSB Topological Iron-Based Superconductor}

\author{Wenyao Liu}
\thanks{These authors contributed equally to this work.}
\affiliation{Department of Physics, Boston College, Chestnut Hill, MA 02467, USA}

\author{Gabriel Natale}
\thanks{These authors contributed equally to this work.}
\affiliation{Department of Physics, Boston College, Chestnut Hill, MA 02467, USA}

\author{Camron Farhang}
\affiliation{Department of Physics and Astronomy, University of California, Irvine, CA 92697, USA}

\author{Michael Geiwitz}
\affiliation{Department of Physics, Boston College, Chestnut Hill, MA 02467, USA}

\author{Qishuo Tan}
\affiliation{Department of Chemistry, Boston University, Boston, MA 02215, USA}

\author{Xingyao Guo}
\affiliation{Department of Physics, Hong Kong University of Science and Technology, Clear Water Bay, Hong Kong, China}

\author{Mason Gray}
\affiliation{Department of Physics, Boston College, Chestnut Hill, MA 02467, USA}

\author{Vincent Lamberti}
\affiliation{Department of Physics, Boston College, Chestnut Hill, MA 02467, USA}

\author{Jazzmin Victorin}
\affiliation{Department of Physics, Boston College, Chestnut Hill, MA 02467, USA}

\author{Huairuo Zhang}
\affiliation{Materials Science and Engineering Division, National Institute of Standards and Technology, Gaithersburg, MD 20899, USA}
\affiliation{Theiss Research, Inc., La Jolla, CA 92037, USA}

\author{James L. Hart}
\affiliation{Department of Materials Science and Engineering, Cornell University, Ithaca, USA}

\author{Vsevolod Belosevich}
\affiliation{Department of Physics, Boston College, Chestnut Hill, MA 02467, USA}

\author{Xi Ling}
\affiliation{Department of Chemistry, Boston University, Boston, MA 02215, USA}
\affiliation{Division of Materials Science and Engineering, Boston University, 15 St. Mary’s Street, Boston, MA 02215, USA}
\affiliation{The Photonics Center, Boston University, 8 St. Mary’s Street, Boston, MA 02215, USA}

\author{Qiong Ma}
\affiliation{Department of Physics, Boston College, Chestnut Hill, MA 02467, USA}

\author{Wan Kyu Park}
\affiliation{National High Magnetic Field Laboratory, Florida State University, FL 32310, USA}

\author{Kenji Watanabe}
\affiliation{Research Center for Electronic and Optical Materials, National Institute for Materials Science, 1-1 Namiki, Tsukuba 305-0044, Japan}

\author{Takashi Taniguchi}
\affiliation{Research Center for Materials Nanoarchitectonics, National Institute for Materials Science, 1-1 Namiki, Tsukuba 305-0044, Japan}

\author{Judy J. Cha}
\affiliation{Department of Materials Science and Engineering, Cornell University, Ithaca, USA}

\author{Albert V. Davydov}
\affiliation{Materials Science and Engineering Division, National Institute of Standards and Technology, Gaithersburg, MD 20899, USA}

\author{Kin Chung Fong}
\affiliation{Department of Electrical and Computer Engineering, Northeastern University, Boston, MA 02115, USA}
\affiliation{Department of Physics, Northeastern University, Boston, MA 02115, USA}
\affiliation{Quantum Materials and Sensing Institute, Northeastern University, Burlington, MA}

\author{Ethan Arnault}
\affiliation{Department of Electrical Engineering and Computer Science, Massachusetts Institute of Technology, Cambridge, MA 02139, USA}

\author{Genda Gu}
\affiliation{Condensed Matter Physics and Materials Science, Brookhaven National Laboratory (BNL), Upton, NY 11973, USA}

\author{Rui-Xing Zhang}
\affiliation{Department of Physics and Astronomy, University of Tennessee, Knoxville, TN 37996, USA}
\affiliation{Department of Materials Science and Engineering, University of Tennessee, Knoxville, TN 37996, USA}

\author{Enrico Rossi}
\affiliation{Department of Physics, William \& Mary, Williamsburg, VA 23187, USA}

\author{Jing Xia}
\affiliation{Department of Physics and Astronomy, University of California, Irvine, CA 92697, USA}

\author{Kenneth S. Burch}
\email{Corresponding author: ks.burch@bc.edu}
\affiliation{Department of Physics, Boston College, Chestnut Hill, MA 02467, USA}

\date{\today}

\begin{abstract}
The interplay of superconductivity, band topology, and spontaneous time-reversal-symmetry breaking (TRSB) is expected to enable topological superconducting boundary states. Intriguingly, the iron-chalcogenide superconductor \ch{FeTe_{0.55}Se_{0.45}} exhibits spontaneous magnetization in the superconducting state, motivating its use as a single-material platform in which three ingredients coexist. Here we report evidence for a nontrivial, boundary-mediated superconducting transport response in exfoliated \ch{Fe(Te,Se)} devices, exhibiting long-range nonlocality and strongly suppressed thermal broadening. Polar Kerr measurements establish that TRSB emerges below $T_{\mathrm{Kerr}}<T_c$ and coexists with superconductivity across multiple Fe(Te,Se) compositions studied here, providing an independent symmetry-breaking scale that can be directly compared with transport. To reliably access boundary transport, we engineer crystallographically sharp, continuous edges and implement side-surface-dominant contacts, as validated by contact-size analysis and cross-sectional imaging. Under these conditions, topological \ch{FeTe_{0.55}Se_{0.45}} exhibits an anomalous conductance plateau that is absent in topologically trivial \ch{FeTe_{0.40}Se_{0.60}} and \ch{Fe_{1.02}Te_{0.55}Se_{0.45}} under comparable measurements. This plateau requires both the source and drain to be attached to uninterrupted sharp edges, persists over micrometer-scale separations far exceeding the bulk coherence length, has markedly reduced thermal decoherence, and collapses when the drain is moved to the top surface. Its temperature evolution is governed by TRSB: the plateau shows minimal thermal broadening below $T_{\mathrm{Kerr}}^{*}$ and disappears near $T_{\mathrm{Kerr}}$ rather than $T_c$. Collectively, the doping selectivity, TRSB correlation, edge-geometry requirement, and long-range nonlocal nature establish experimentally grounded criteria for identifying boundary-mediated superconducting transport in \ch{FeTe_{0.55}Se_{0.45}}, and motivate further phase-sensitive and theoretical work to determine its microscopic origin.
\end{abstract}

\maketitle


\section*{Introduction}
The realization of superconductivity, band topology, and magnetism within a single quantum material has long been pursued.\cite{fu2008superconducting,lutchyn2010majorana,qi2011topological,linder2015superconducting,cai2023superconductor,yi2024interface} Their coexistence enables new superconducting states and transport responses, including finite-momentum Cooper pairing,\cite{wan2023orbital} nonreciprocal supercurrents\cite{yuan2022supercurrent,chen2024intrinsic}, and protected boundary modes.\cite{kallin2012chiral, alicea2012new,yazdani2023hunting,hasan2015topological} However, transport and tunneling measurements in candidate topological superconductors have proven difficult to interpret. Unlike topological insulators or semimetals, where boundary-state transport can in principle be separated from an insulating or weakly conducting bulk, a topological superconductor contains a bulk superconducting condensate. As a result, bulk supercurrents, quasiparticle leakage, and Andreev processes can obscure boundary contributions, making standard boundary-transport strategies difficult to apply directly.\cite{kim2024edge} In addition, local zero-bias conductance anomalies, although widely used as probes of low-energy boundary excitations, are not unique signatures of topology or Majorana quasiparticles; similar features can arise from trivial Andreev bound states (ABS), disorder, contact inhomogeneity, heating, or bulk-mediated processes.\cite{kayyalha2020absence,sarma2021disorder,hui2015bulk,gifford2016zero} Furthermore, additional current at the edge could arise from inhomogeneity and screening current, rather than indicating a truly protected, nonlocal phenomenon.  

A particularly compelling prospect is that topological superconducting phases can uniquely exhibit long-range, nonlocal superconducting transport responses that are distinct from ordinary local Andreev and bulk quasiparticle conduction.\cite{serban2010domain,he2014correlated,zhang2017quantum,ikegaya2019anomalous} Demonstrating such responses experimentally requires more than observing a local zero-bias feature. Nevertheless, transport-based demonstrations of such protected nonlocal responses remain challenging. It requires a feasible material platform combining superconductivity, topology, and symmetry breaking, together with precise control over edge morphology, contact quality, and measurement geometry.\cite{hui2015bulk,kayyalha2020absence,sarma2021disorder,gifford2016zero,ikegaya2019anomalous} In particular, a stringent transport test should distinguish a local response from a boundary-mediated signal that depends on the continuity of the edge pathway.

In this context, iron-chalcogenide superconductors \ch{FeTe_{1-x}Se_{x}} [\ch{Fe(Te,Se)}] have emerged as a particularly attractive platform, as for suitable Te/Se ratios they combine robust superconductivity with a topologically nontrivial electronic structure near the Fermi level.\cite{wang2015topological,zhang2018observation,li2021electronic, Lohani.2019} Experiments on \ch{FeTe_{0.55}Se_{0.45}} have reported vortex-core zero-energy bound states and boundary-associated anomalies, pointing to modes resulting from topological superconductivity.\cite{zhang2018observation,choi2022emergent,hatefipour2022a,wang2020evidence,gray2019evidence} Beyond band topology and superconductivity, a key recent development is that \ch{Fe(Te,Se)} also exhibits time-reversal-symmetry breaking (TRSB) within the superconducting state. Experiments using different methods and samples revealed spontaneous magnetization in superconducting states of Fe(Te,Se).\cite{zaki2021time,mclaughlin2021strong,matsuura2023two} Notably, the matching temperature dependence observed by Sagnac magneto-optical Kerr effect (SMOKE) and $\mu$SR shows that TRSB occurs in the bulk of the samples.\cite{farhang2023revealing,roppongi2025topology} This additional symmetry breaking provides a natural setting to seek new topological superconducting phases, motivating edge-contact transport for the identification of protected nonlocal responses.\cite{meng2012weyl,sau2012topologically,yazdani2023hunting,wu2021topological,hu2024dislocation,roppongi2025topology}

Here, we report an experimental methodology for identifying boundary-mediated superconducting transport in topological \ch{Fe(Te,Se)} flakes. We utilize a Cleanroom-in-glovebox fabrication process, where samples are isolated from the ambient environment throughout the entire process (see Methods), to prepare high-quality edge-contact devices on exfoliated \ch{Fe(Te,Se)} flakes with crystallographically defined terminal edges. Our strategy is to establish a hierarchy of experimentally accessible criteria distinguishing the boundary-mediated superconducting transport response from conventional Andreev physics, bulk-mediated transport, and fabrication-related artifacts, and test these criteria systematically.

First, in Sec.~\ref{sec:Results1}, we demonstrate that by controlling both the composition and the crystallographic edge morphology, topological \ch{FeTe_{0.55}Se_{0.45}} exhibits two distinct edge-associated responses in transport spectroscopy: a previously unobserved conductance plateau that appears only when contacts are fabricated on sharp, continuous edges, and a zero-bias conductance peak (ZBCP) on rough or stepped edges, consistent with our previous work.\cite{gray2019evidence} In contrast, topologically trivial \ch{FeTe_{0.40}Se_{0.60}} and \ch{Fe_{1.02}Te_{0.55}Se_{0.45}}\cite{li2021electronic} display only conventional Andreev-reflection (AR) spectra under the same conditions.\cite{daghero2010probing,tang2019quasi} This combined dependence on composition and edge morphology indicates that the plateau is not a generic contact or edge-roughness effect.

In Sec.~\ref{sec:Results2}, we constrain trivial explanations by combining transport and SMOKE measurements, establishing that the plateau is not attributable to degraded bulk superconductivity\cite{zalic2019fete}, or compositional inhomogeneity near the edge. In Sec.~\ref{sec:Results3}, we further show that the plateau is insensitive to the measurement implementation: it persists under both DC and AC excitation and across two- and three-terminal configurations, disfavoring circuit-induced or critical-current-related artifacts. We also apply an effective contact-size analysis supported by cross-sectional STEM, which, when combined with measurements on devices with nanoscale-wide contacts, strongly argues against non-ballistic contact artifacts.\cite{sheet2004role,gifford2016zero} 

Having constrained these local and contact-related explanations, in Sec.~\ref{sec:Results4} we demonstrate that the plateau exhibits a pronounced nonlocal response. It requires a continuous edge pathway connecting the source and drain over micrometer-scale distances, far exceeding the bulk coherence length. These features are incompatible with conventional bulk-mediated nonlocal mechanisms in superconductors.\cite{russo2005experimental,cadden2006nonlocal,morten2006circuit,galluzzi2019transport} Finally, in Sec.~\ref{sec:Results5}, temperature-dependent measurements reveal that the edge-associated transport in the topological composition is governed by the TRSB state. 
The plateau exhibits strongly suppressed thermal broadening below $T_{Kerr}^*$ and disappears near $T_{Kerr}$ rather than at $T_c$, in sharp contrast to bulk-state spectra that track the superconducting gap and persist up to $T_c$.\cite{park2010strong,tang2019quasi} Notably, the plateau's width ($\sim 0.8$ mV) is also comparable to the energy scale associated with $T_{Kerr}$, providing an additional indication that the plateau is tied to the TRSB state. This distinct temperature scale further separates the plateau from conventional Andreev or trivial edge-bound-state physics, which should primarily follow the superconducting order parameter.

Taken together, our results establish previously unobserved nonlocal and topologically-associated signatures mediated by edge modes resulting from the combination of topology, superconductivity, and TRSB in \ch{FeTe_{0.55}Se_{0.45}}. 
Even in the absence of a complete microscopic theory, such an experimentally grounded framework is essential for isolating boundary-mediated superconducting transport in a single-material platform. This work therefore addresses a central methodological challenge in the transport study of topological superconductors and motivates further microscopic theory and phase-sensitive experiments to determine the underlying boundary excitations.

\section{\label{sec:Results1}Edge-geometry-defined transport responses}

\begin{figure} [htb]
    \centering
    \includegraphics[width=1\linewidth]{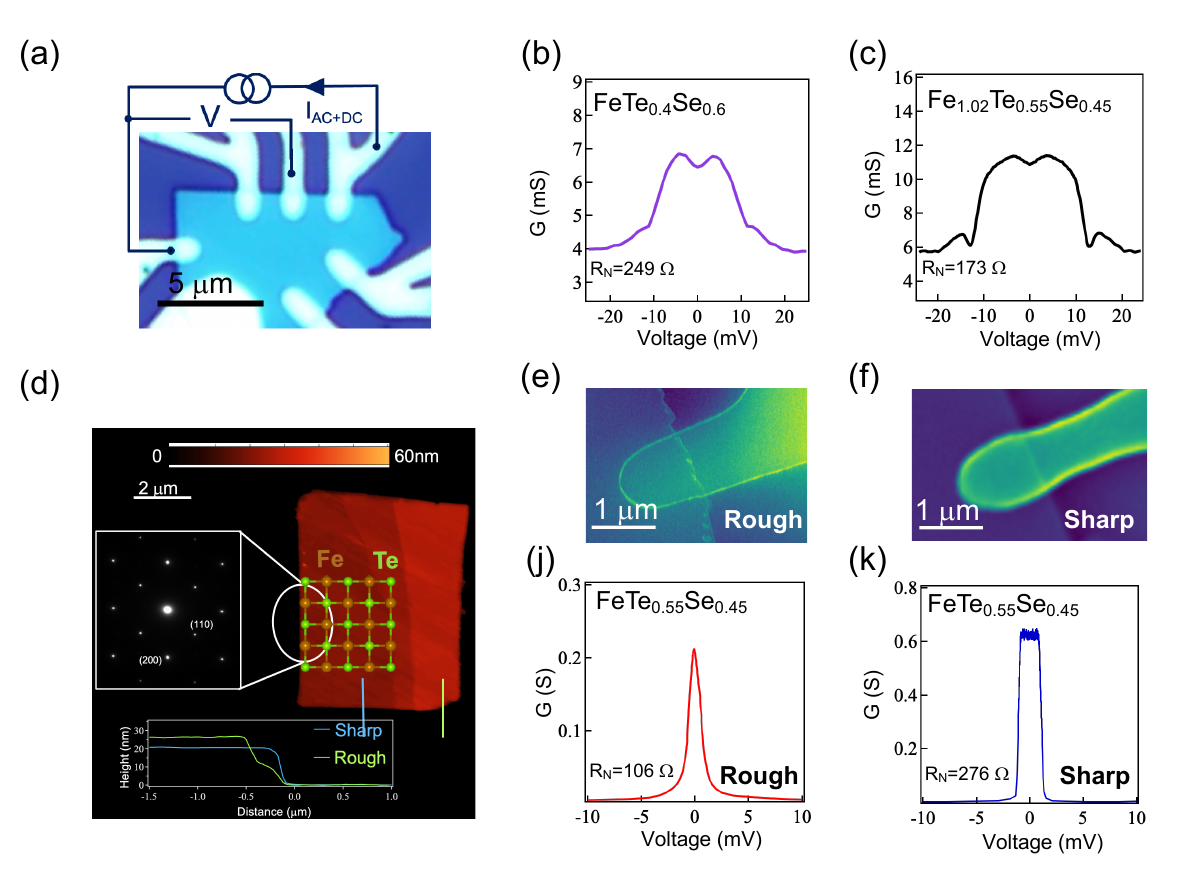}
    \caption{\label{fig:introEXP} Edge-geometry-dependent conductance spectra in Fe(Te,Se) devices. (a) Optical image of a representative exfoliated Fe(Te,Se) device, together with the measurement configuration. Differential conductance ($G=\frac{dI}{dV}$)is measured using an AC+DC bias applied to the source electrode while the drain is grounded. (b,c) Differential conductance spectra measured at 1.4 K on topologically trivial Fe(Te,Se) devices, including (b)\ch{FeTe_{0.40}Se_{0.60}} and (c)\ch{Fe_{1.02}Te_{0.55}Se_{0.45}}. (d) AFM topography of an exfoliated Fe(Te,Se) flake. The flake exhibits both a straight, crystallographically aligned sharp edge (Line 1) and a rough edge containing steps and irregularities (Line 2). Inset: selected-area electron diffraction pattern confirming the crystalline orientation. (e,f) False-color scanning electron microscopy (SEM) images showing electrodes contacting a rough edge (e) and a sharp, continuous edge (f) of \ch{FeTe_{0.55}Se_{0.45}} flakes, respectively. (j) Differential conductance measured at 1.4 K on a \ch{FeTe_{0.55}Se_{0.45}} device with electrodes contacting a rough edge, exhibiting a zero-bias conductance peak. (k) Differential conductance measured at 1.4 K on a \ch{FeTe_{0.55}Se_{0.45}} device with electrodes contacting a sharp, continuous edge, showing a bias-independent conductance plateau within a finite voltage window. Corresponding normal-state contact resistances $R_N$ are indicated in each panel.
}
\end{figure}

To establish an experimental criterion for distinguishing boundary-mediated transport of topological origin from bulk superconducting responses, we first examine how the differential conductance spectra [see the configuration in Fig.~\ref{fig:introEXP}(a)] depend on the material composition and edge morphology of Fe(Te,Se) devices (Fig.~\ref{fig:introEXP}). Here we are motivated by the importance of a well-defined surface normal to prevent unwanted signals in point-contact Andreev spectroscopy studies.\cite{deutscher2005andreev} Specifically, we benefit from exfoliated Fe(Te,Se) flakes' naturally occurring edges with distinct morphologies set by their cleavage behavior. Owing to the tetragonal crystal structure [Fig.~\ref{fig:introEXP}(d)], exfoliation commonly produces edges aligned along the [100] or [010] directions, as characterized by selected-area electron diffraction [inset of Fig.~\ref{fig:introEXP}(d)]. Such crystallographic edges form well-defined $90^{\circ}$ corners and provide straight, continuous side-surface terminations (as sharp edges). A continuous boundary of this type is, in principle, an essential geometric condition for observing transport responses mediated by propagating edge modes, since steps or discontinuities can interrupt the boundary pathway.\cite{ikegaya2019anomalous,zhao2020interference,he2014correlated,zhang2017quantum} 
In contrast, edges that deviate from the [100]/[010] directions are typically rough and stepped. Atomic force microscopy (AFM) line profiles across rough edges reveal additional steps and nanometer-scale height variations [Fig.~\ref{fig:introEXP}(d)], indicating multiple local facets and discontinuities along the side surface. Electrodes contacting these rough edges can therefore couple to a geometrically fragmented boundary rather than to a single continuous, straight interface. As shown below, while the conductance spectra of topologically trivial devices show no qualitative dependence on edge morphology, topological \ch{FeTe_{0.55}Se_{0.45}} devices exhibit qualitatively distinct edge-associated responses that sensitively depend on whether the contacted boundary is sharp or rough.

We begin by taking reference spectra using Fe(Te,Se) compositions that are topologically trivial but exhibit superconductivity and magnetic transitions (as shown later) similar to topologically nontrivial \ch{FeTe_{0.55}Se_{0.45}}.\cite{li2021electronic,farhang2023revealing,roppongi2025topology} Devices fabricated from \ch{FeTe_{0.40}Se_{0.60}} and \ch{Fe_{1.02}Te_{0.55}Se_{0.45}} were prepared from exfoliated flakes with a thickness around 30 nm. For both compositions, differential conductance spectra (see Methods) measured at 1.4 K [Fig.~\ref{fig:introEXP}(b,c)] display features characteristic of conventional AR process\cite{daghero2011directional,deutscher2005andreev}, namely, a conductance enhancement within a finite bias window and symmetric coherence-peak-like structures located around 4 mV [Fig.~\ref{fig:introEXP}(b,c)]. These spectra on trivial Fe(Te,Se) devices are in good agreement with those previously reported on Fe(Te,Se) thin films and bulk crystals\cite{daghero2010probing,tang2019quasi}, where the superconducting gap is on the order of several meV, and the spectra can be well described by the Blonder–Tinkham–Klapwijk (BTK) framework (see Supplemental Materials)\cite{daghero2010probing,tang2019quasi}. Importantly, the spectra of these trivial devices show no qualitative dependence on whether the contacted boundary is sharp or rough (see Supplemental Materials), thereby providing a robust baseline for comparison.

Next, we focus on devices fabricated from topological \ch{FeTe_{0.55}Se_{0.45}} flakes. When electrodes are placed on rough/stepped edges of \ch{FeTe_{0.55}Se_{0.45}} flakes, identified by scanning electron microscopy [SEM, Fig.~\ref{fig:introEXP}(e)], the differential conductance shows a pronounced zero-bias conductance peak [Fig.~\ref{fig:introEXP}(j)], consistent with earlier reports in edge- or surface-contact geometries without explicit control of edge morphology.\cite{gray2019evidence} Significantly, a different response emerges when electrodes are placed on sharp, straight, and continuous flake edges [see SEM in Fig.~\ref{fig:introEXP}(f)]. In this case, the conductance spectrum exhibits a bias-independent conductance plateau [Fig.~\ref{fig:introEXP}(k)]. The plateau feature only appears within a narrow bias window of approximately 1 meV (well inside superconducting gap $\Delta_{SC}\approx 4~meV$)\cite{park2010strong,tang2019quasi,wang2018evidence}, and displays a much larger sub-gap conductance enhancement relative to its normal-state background compared to spectra in \ch{Fe_{1.02}Te_{0.55}Se_{0.45}} [Fig.~\ref{fig:introEXP}(b)], \ch{FeTe_{0.40}Se_{0.60}} [Fig.~\ref{fig:introEXP}(c)], or other superconductors\cite{daghero2010probing,daghero2012strong}. 

We note that the transition from a ZBCP to a conductance plateau in \ch{FeTe_{0.55}Se_{0.45}} flakes is achieved solely by changing the contacted edge morphology, without altering the bulk material composition, device thickness, measurement temperature, or normal-state resistance (as discussed later). Notably, to move beyond a qualitative morphological description, we established an empirical quantitative threshold governing this plateau-to-ZBCP transport transition. The conductance plateau is reproducibly observed across fifteen independent contacts on sharp edges of topological \ch{FeTe_{0.55}Se_{0.45}} devices. Statistical analysis across these devices (see Fig. S9 in Supplemental Material) reveals that the robust conductance plateau is exclusively resolved on 'sharp' edges characterized by a macroscopic in-plane tortuosity, 
i.e., the ratio of the actual traced length of the contacted edge to the shortest straight-line distance,  $\tau \lesssim 1.015$,  and a microscopic edge transition-width-to-thickness ratio ($W/H$) below $10$.\cite{ghanbarian2013tortuosity,vandenberghe2017imperfect} Exceeding these geometric thresholds introduces a terraced multi-step morphology that disrupts the continuous boundary, invariably driving the transport signature into a ZBCP.

Moreover, this strong dependence on edge geometry is absent in topologically trivial \ch{Fe(Te,Se)} devices (see Supplemental Material), indicating that the observed plateau is not a generic consequence of superconductivity or contact quality. Taken together, we observed a clear experimental distinction between three transport regimes: conventional AR-like spectra in topologically trivial \ch{Fe(Te,Se)}, a ZBCP associated with rough edges in \ch{FeTe_{0.55}Se_{0.45}}, and a robust bias-independent conductance plateau observed only when sharp edges of \ch{FeTe_{0.55}Se_{0.45}} are contacted.

\section{\label{sec:Results2}Transport and magnetic characterization}

\begin{figure} [htb]
    \centering
    \includegraphics[width=1\linewidth]{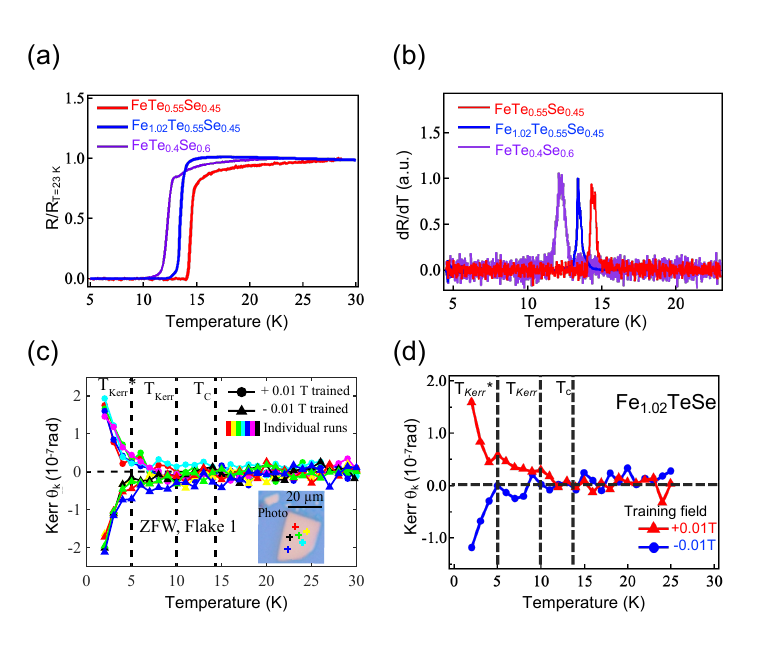}
    \caption{\label{fig:CharaFTS} Transport and magnetic characterization, and effective contact-size analysis of Fe(Te,Se) devices. (a) Temperature-dependent four-terminal resistance measured on exfoliated \ch{FeTe_{0.55}Se_{0.45}}, \ch{Fe_{1.02}Te_{0.55}Se_{0.45}}, and \ch{FeTe_{0.40}Se_{0.60}} devices. (b) Temperature derivative $dR/dT$ extracted from the data in (a). (c) SMOKE measurements during zero-magnetic-field warming on an Fe(Te,Se) flake after cooling in $+0.01$ T (circles) and $-0.01$ T (triangles) training fields. Different colors represent measurements at various locations. Inset: photo of the Fe(Te,Se) flake, with measurement locations marked with "+". (d) Temperature dependence of the polar Kerr rotation $\theta_{K}$ measured on \ch{Fe_{1.02}Te_{0.55}Se_{0.45}}.  
    }
\end{figure}

The anomalous conductance plateau observed in \ch{FeTe_{0.55}Se_{0.45}} devices suggests the presence of additional electronic states associated with the sample edges. However, such features could in principle still arise from trivial origins, including contact artifacts, edge-induced inhomogeneity, or differences in bulk superconducting or magnetic properties among Fe(Te,Se) compositions. To explore and exclude these possibilities, we performed systematic transport, magnetic, and structural characterizations on Fe(Te,Se) devices (Fig.~\ref{fig:CharaFTS}).

We first examine the bulk superconducting properties using four-terminal resistance measurements. As shown in Fig.~\ref{fig:CharaFTS}(a), exfoliated devices fabricated from \ch{FeTe_{0.55}Se_{0.45}}, \ch{Fe_{1.02}Te_{0.55}Se_{0.45}}, and \ch{FeTe_{0.40}Se_{0.60}} exhibit comparable superconducting transitions ($T_c$). To further quantify the transition sharpness, we analyze the temperature derivative of the resistance, $dR/dT$ [Fig.~\ref{fig:CharaFTS}(b)]. All three compositions display similar transitions, with $T_{c} \approx 14.2$ K for \ch{FeTe_{0.55}Se_{0.45}}, $\approx 13.5$ K for \ch{Fe_{1.02}Te_{0.55}Se_{0.45}}, and $\approx 12.5$ K for \ch{FeTe_{0.40}Se_{0.60}}. The similarity in transition widths and $T_c$ indicates that the superconducting transitions are comparable and occur over a narrow temperature window. Most crucially, there is no evidence for strongly broadened or multi-step transitions that would suggest pronounced phase separation or macroscopic inhomogeneity. Although the normal-state resistivity backgrounds differ among compositions (e.g., \ch{Fe_{1.02}Te_{0.55}Se_{0.45}} shows a more insulating-like $R(T)$ above $T_c$), the sharp onset of superconductivity in all cases establishes that the distinct low-temperature differential-conductance responses discussed below are not simply a consequence of a degraded or highly inhomogeneous superconducting transition.

Beyond the superconductivity, these three dopings could exhibit substantial differences in TRSB, as found in previous works on Fe(Te,Se) \cite{mclaughlin2021strong,farhang2023revealing,roppongi2025topology,matsuura2023two}. Thus, we characterized the finite magnetization of our exfoliated samples using Sagnac magneto-optical Kerr effect (SMOKE) measurements, where the polar-Kerr signal (\(\theta_{K}\)) is proportional to the perpendicular component of the magnetization \(M_{\perp}\)\cite{argyres_theory_1955}. As shown in Fig.~\ref{fig:CharaFTS}(c), $\theta_{K}$ is measured on topological \ch{FeTe_{0.55}Se_{0.45}} flakes during zero-field warming after cooling in small positive and negative training fields (0.01 T). The training field dependence of $\theta_{K}$ reveals the onset of spontaneous magnetization at $T_{Kerr} \approx 10$ K, which is highly consistent with recent $\mu$SR measurements on Fe(Te,Se) \cite{roppongi2025topology} and well below $T_{c}$. In addition, a rapid enhancement of $\theta_{K}$ is observed below $T_{Kerr}^* \approx 5$ K, which may result from re-orientation of the magnetic easy-axis (see Supplemental Materials). Here, we also found that $\theta_{K}$ is identical regardless of the training field's magnitude (see Supplemental Materials), which rules out pinned vortices as its origin. Furthermore, the transition temperature was universal on numerous flakes under different conditions: as exfoliated, in devices, and with various thicknesses and protective layers (see Supplemental Materials). Similar Kerr measurements also performed on non-topological \ch{Fe_{1.02}Te_{0.55}Se_{0.45}} devices [Fig.~\ref{fig:CharaFTS}(d)] demonstrate a comparable magnetic transition. Notably, the Kerr signal is reproducible across different spatial locations on the same flake [Fig.~\ref{fig:CharaFTS}(c)], indicating uniform magnetic ordering.

In addition, we also examine the possibility of edge-induced compositional inhomogeneity. Energy-dispersive X-ray spectroscopy (EDX) mapping was performed on representative exfoliated \ch{FeTe_{0.55}Se_{0.45}} flakes (see Supplemental Materials). The elemental maps for Fe, Te, and Se are spatially uniform along the edge as well as across the interior of the flake. Crucially, no evidence of phase separation or compositional variation is observed at the rough or sharp edges, ruling out edge-induced inhomogeneity as the origin of the observed edge-dependent conductance responses.

\section{\label{sec:Results3}Independence of the plateau from the circuit-configuration, and effective contact size}

\begin{figure*} [htb]
    \centering
    \includegraphics[width=0.85\linewidth]{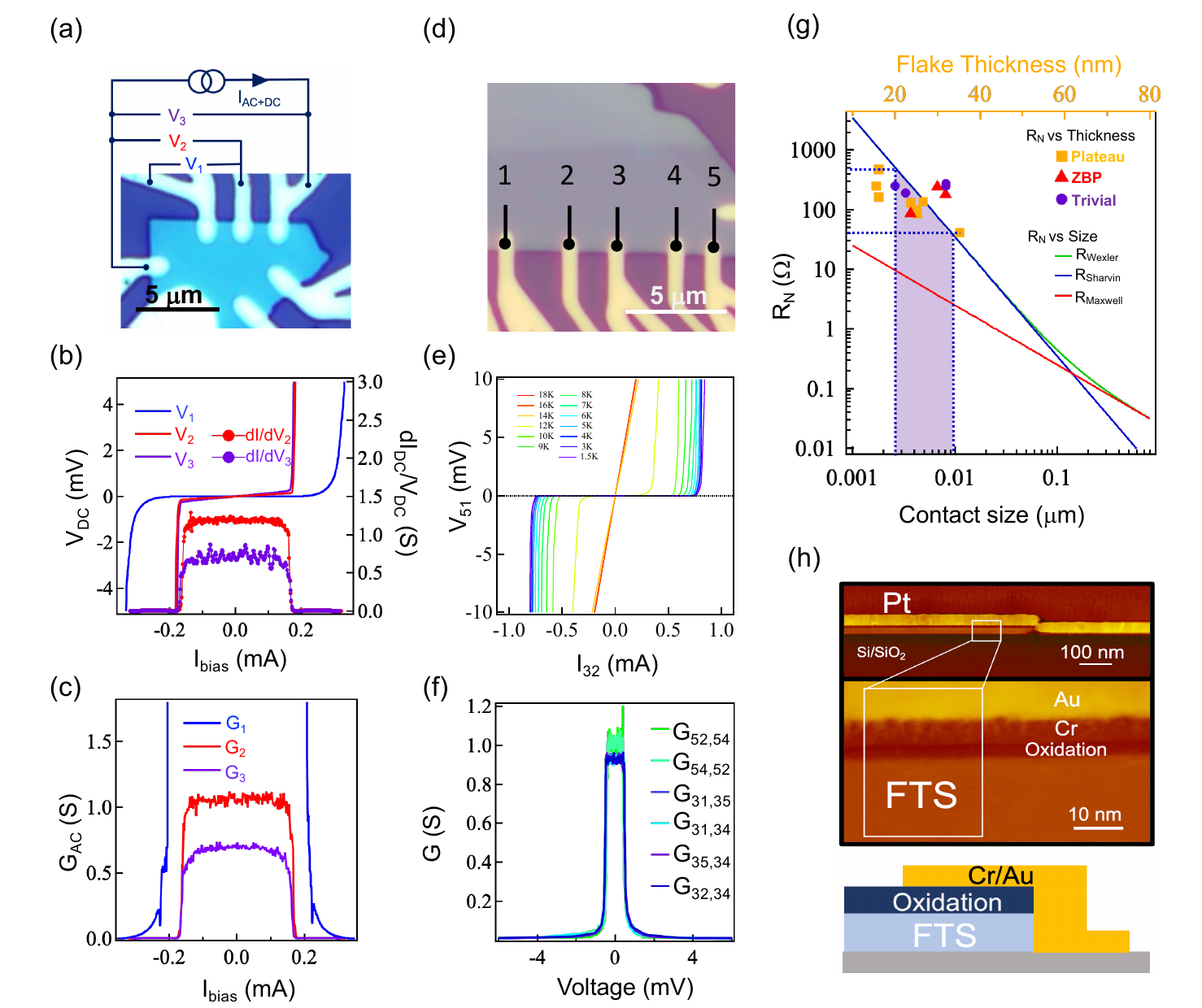}
    \caption{\label{fig:CircuitFTS} Circuit-configuration-dependent transport measurement. (a) Optical image of \ch{FeTe_{0.55}Se_{0.45}} device with multiple voltage probes, illustrating the configurations used for two-point, three-point, and four-point measurements. (b) DC current–voltage (I–V) characteristics measured at 1.4 K using two-, three-, and four-point configurations. The corresponding conductance in three-point and four-point configurations is extracted by numerical differentiation. (c) Corresponding AC differential conductance spectra measured simultaneously at different voltage probes in the Device A. (d) \ch{FeTe_{0.55}Se_{0.45}} device B with five electrical contacts (1-5) labeled for multi-terminal measurements. (e) Temperature-dependent four-point DC transport measurements performed on this device, showing the evolution of the bulk superconducting response upon cooling. (f) Three-point differential conductance spectra measured on Device B using different current injection directions and voltage probe combinations, where $G_{ij,hk}=\mathrm{d}I_{ij}/\mathrm{d}V_{hk}$. The conductance plateau exhibits consistent height and spectral features across all measurement configurations. (g) Normal-state contact resistance $R_{N}$ plotted as a function of the effective contact size estimated using the Sharvin resistance, Maxwell resistance, and the Wexler interpolation. The shaded region indicates the range of effective contact sizes extracted from the measured $R_{N}$. Symbols denote contacts associated with conductance plateaus, zero-bias peaks, and conventional Andreev-feature spectra. (h) Schematic illustration of the edge-contact geometry and representative device structure. Cross-sectional ADF-STEM images show that the top surface of the electrode is electrically isolated by a residual oxidized layer, resulting in predominantly edge-contacted transport. 
    }
\end{figure*}

Having examined bulk superconductivity, magnetic order(Fig.~\ref{fig:CharaFTS}), and compositional uniformity (Supplemental Materials), we next test whether the conductance plateau could arise from specific measurement configurations or circuit-related artifacts. To this end, we systematically compare transport measurements performed using different probing configurations and contact geometries (Fig.~\ref{fig:CircuitFTS}). We first focus on the \ch{FeTe_{0.55}Se_{0.45}} device A [Fig.~\ref{fig:CircuitFTS}(a)], in which four-terminal ($V_1$), three-terminal ($V_2$), and two-terminal ($V_3$) pure DC transport measurements are performed using the same set of source-drain contacts. As shown in Fig.~\ref{fig:CircuitFTS}(b), the measured current–voltage characteristics change as expected from the different configurations. Specifically, the four-terminal measurement exhibits a well-defined zero-resistance state below $T_{c}$ when the applied current is below the critical current. In contrast, the two-terminal and three-terminal measurements capture a finite voltage signal with a constant slope, even when the applied current is well below the critical current. The corresponding conductance spectra obtained by numerically differentiating the current-voltage results (dI$_{DC}$/dV$_{DC}$), display plateau spectra with the conductance offset from the additional contact resistance of the drain. 

We further directly compare these pure DC transport results with AC differential conductance spectroscopy using the same device and contact configuration. As shown in Fig.~\ref{fig:CircuitFTS}(c), the bias-independent conductance plateau observed in DC measurements is reproduced in the AC differential conductance spectra, with consistent magnitude and voltage range. This agreement demonstrates that the plateau is not an artifact of DC current biasing, such as local Joule heating or critical-current-induced switching\cite{daghero2010probing,duif1989point}, but rather reflects an intrinsic low-energy transport response.

This conclusion is further substantiated by the robustness of the plateau response to changes in current injection direction and electrode configuration. As shown in Fig.~\ref{fig:CircuitFTS}(d), a second \ch{FeTe_{0.55}Se_{0.45}} flake (Device B) was fabricated with five terminals along the same sharp edge. Figure~\ref{fig:CircuitFTS}(e) displays temperature-dependent four-terminal DC transport measured along this edge, confirming robust edge superconductivity with T$_c$ = 14~K. Figure~\ref{fig:CircuitFTS}(f) shows three-terminal measurements performed on this edge, where the differential conductance $G_{ij,hk}=\mathrm{d}I_{ij}/\mathrm{d}V_{hk}$ is extracted using different combinations of current and voltage leads along the same edge. In all configurations where both current and voltage probes couple to the sharp edge, a conductance plateau with nearly identical height and voltage width is observed. The reproducibility of the plateau under these variations confirms that the response does not depend on a specific lead assignment, current direction, or accidental circuit asymmetry. 

As a last check, we assess whether the observed ZBCP and plateau features could arise from artifacts associated with large or non-ideal contacts. In three-terminal point-contact measurements, zero-bias anomalies can emerge from inelastic scattering or non-spectroscopic effects\cite{sheet2004role,gifford2016zero}, where the relevant length scale in such measurements is the effective size of the electronic interface.\cite{daghero2011directional,deutscher2005andreev} Although this effective contact size cannot be measured directly\cite{chen2010pronounced,sheet2004role}, it is routinely estimated from the normal-state contact resistance $R_{N}$ using standard point-contact models, including the Sharvin (ballistic), Maxwell (diffusive), and Wexler (interpolation) formulations.\cite{daghero2011directional,deutscher2005andreev} We calculated the expected $R_{N}$ over a wide range of effective contact sizes using the Fermi wave vector of Au and the resistivity of the electrodes (see Supplemental Materials), and compare these values with the experimentally measured $R_{N}$ for contacts exhibiting conductance plateaus, ZBCPs, and usual Andreev spectra [Fig.~\ref{fig:CircuitFTS}(g)]. The measured $R_{N}$ values correspond to effective contact sizes of approximately 10–100 nm, a regime where the resistance is dominated by the Sharvin contribution [blue line in Fig.~\ref{fig:CircuitFTS}(g)]. By contrast, assuming a purely diffusive (Maxwell-dominated) contact would require an unrealistically small size (far below 1 nm), well below the mean free path of electrons in the metallic electrodes, which is inconsistent with the model's assumption. 

We further corroborate this conclusion using devices fabricated via Thermal Scanning Probe Lithography (t-SPL), in which the electrode width and contact footprint are reduced from $\sim$0.6 $\mu$m (photolithography-fabricated devices) to the nanometer scale ($\sim$200 nm, see Supplementary Material). Despite the dramatically smaller contact area, these devices reproduce the same edge-associated plateau on sharp edges and ZBCP on rough/stepped edges. This demonstrates that the plateau/ZBCP phenomenology does not rely on an anomalously large contact area or a specific lithographic methodology; instead, it reflects an intrinsic edge-coupled transport response.

Furthermore, the length scale of effective contact sizes is comparable to the thicknesses of the exfoliated Fe(Te,Se) flakes used in this study (20–40 nm) and does not depend on composition. These results indicate that our methodology yields current injection predominantly through the side surface rather than the top surface. This conclusion is independently supported by cross-sectional ADF-STEM measurements [Fig.~\ref{fig:CircuitFTS}(h) and Supplemental Materials], which reveal a residual oxidized layer separating the metallic electrode from the top surface of the Fe(Te,Se) flake, effectively suppressing direct top-surface electrical coupling. Here, such oxidation is known to develop rapidly in thin Fe(Te,Se) flakes even under inert conditions \cite{zalic2019fete}. 
To address this challenge, we utilized our 'Cleanroom in a Glovebox' system (see Methods) which avoids exposing freshly exfoliated flakes to ambient conditions throughout the entire sample preparation and device fabrication process.\cite{gray2020cleanroom,gray2019evidence} When combined with optimized Ar plasma treatment that preferentially removes the oxide layer on the side surface prior to metal deposition, this enables reliable electrical contact predominantly along the exposed edge (see Methods). Thus, the contact-size analysis and structural characterization demonstrate that the measured conductance primarily originates from side-surface transport rather than from top-surface or electrode-dominated pathways.

\section{\label{sec:Results4}Nonlocal transport conductance}
\begin{figure} [htb]
    \centering
    \includegraphics[width=1\linewidth]{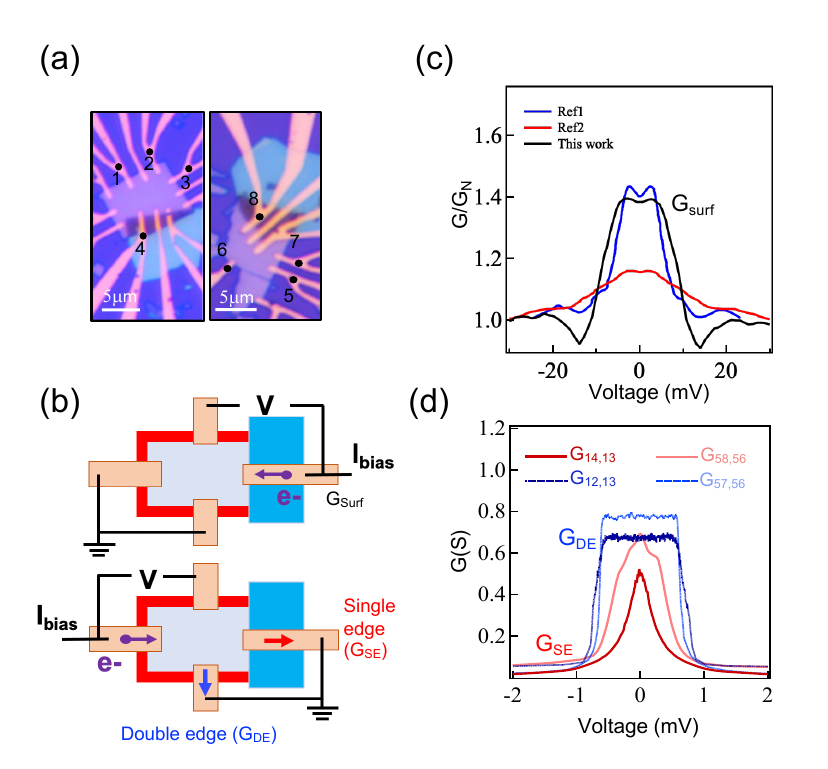}
    \caption{\label{fig:NonlocalFTS} Nonlocal transport response of edge-state conductance in Fe(Te,Se) devices. (a) Top-view images of Fe(Te,Se) devices, showing the electrode layout and contact numbering used for multi-terminal measurements. The Fe(Te,Se) flakes are partly covered by an hBN flake (blue slab). (b) Schematics of the measurement configurations defining double-edge-lead ($G_{DE}$), single-edge-lead ($G_{SE}$), and surface-lead ($G_{Surf}$) configurations. (c) $G_{41,42}$ measured in the surface-contact configuration at 1.4 K. For comparison, FTS surface-contact measurements labeled "ref1" and "ref2" are extracted from previous studies\cite{tang2019quasi,park2010strong}; all spectra are normalized by the normal-state conductance $G_{N}$. (d) Differential conductance $G_{ij,hk}=\mathrm{d}I_{ij}/\mathrm{d}V_{hk}$ measured at 1.4 K in the double-edge-lead and single-edge-lead configurations, where the electrical contacts labeled on the devices shown in (a).
    }
\end{figure}

As mentioned in Sec.~\ref{sec:Results1}, the conductance plateau and ZBCP depend sensitively on the contacted edge morphology and are observed only in topological \ch{FeTe_{0.55}Se_{0.45}} devices. These features, therefore, suggest an origin associated with boundary electronic states in the superconducting phase. To further test this interpretation, we designed experiments that directly probe the requirement for edge contacts in the transport process responsible for the conductance plateau in \ch{FeTe_{0.55}Se_{0.45}} devices. Specifically, we fabricated devices partially covered by thick hexagonal boron nitride (hBN), which selectively blocks access to portions of the top surface and one edge [Fig.~\ref{fig:NonlocalFTS}(a)], enabling controlled comparison between surface-only contacts and sharp-edge contacts within the same device, as well as direct comparison to our trivial Fe(Te,Se) devices and previous point contact experiments \cite{tang2019quasi,park2010strong,daghero2011directional}.

We first examine whether the plateau could arise from superconductivity in the Dirac surface states of \ch{FeTe_{0.55}Se_{0.45}}, namely, the Klein-paradox.\cite{lee2019perfect} To this end, surface-contact measurements are performed where the source current and voltage probes couple exclusively to the top surface (surface-lead configuration; upper panel of Fig.~\ref{fig:NonlocalFTS}(b)). As shown in Fig.~\ref{fig:NonlocalFTS}(c), the differential conductance measured via the surface lead ($G_{Surf}$) exhibits the conventional AR characteristics that is similarly observed in trivial Fe(Te,Se) devices [Fig.~\ref{fig:introEXP}(b,c)], and is consistent with previous surface-contact measurements on \ch{FeTe_{0.55}Se_{0.45}} in Fig.~\ref{fig:NonlocalFTS}(c)\cite{tang2019quasi,park2010strong}. These results reveal that even in the presence of superconducting Dirac-like surface states, transport through the top surface produces ordinary Andreev features. 

Next, to distinguish a purely local tunneling/Andreev process at the injection edge contact from an edge-mediated nonlocal transport process, we exploit a key feature of our multi-terminal geometry: since the superconducting bulk of the Fe(Te,Se) flake is shorted by dissipation-less supercurrent, a local process at the source interface must not depend on where the drain electrode is placed.\cite{zeng2006local} Thus, we compare configurations that keep the source contact unchanged while selectively moving the drain between an edge-coupled and a surface-coupled electrode [lower panel of Fig.~\ref{fig:NonlocalFTS}(b)]. In the single-edge-lead configuration, the drain is displaced from the edge to the top surface, while the source remains edge-coupled. In this configuration, the plateau didn't appear, and the differential conductance develops a pronounced ZBCP, as observed for $G_{14,13}$ in Device~A and $G_{58,56}$ in Device~B [Fig.~\ref{fig:NonlocalFTS}(d)]. 

We next restore the uninterrupted edge pathway between source and drain. In the double-edge-lead configuration, both electrodes couple to the same continuous sharp edge [Fig.~\ref{fig:NonlocalFTS}(b)]. The conductance reverts to a well-defined plateau within $\pm 0.8$~mV, observed as $G_{12,13}$ in Device~A and $G_{57,56}$ in Device~B [Fig.~\ref{fig:NonlocalFTS}(d)], confirming reproducibility across devices. Importantly, this qualitative switching between a plateau and a ZBCP is achieved by relocating only the drain while keeping the source contact and measurement temperature fixed. This rules out explanations based on a particular ``good'' source contact or an idiosyncratic electrode, and instead indicates that the plateau requires coherent coupling between spatially separated contacts along the same continuous edge. Consistent with this interpretation, the plateau is reproduced under source--drain interchange and different current directions (Sec.~\ref{sec:Results3}), and it is observed only when the source and drain share an uninterrupted sharp edge (Sec.~\ref{sec:Results1}); contacting a rough/stepped edge---which effectively breaks edge continuity---suppresses the plateau and yields a ZBCP. Taken together, these results indicate the plateau results from a nonlocal process mediated by an additional edge mode.

Furthermore, we note that the source--drain separation is on the order of micrometers, far exceeding the bulk superconducting coherence length of Fe(Te,Se) ($\xi \sim 6$ nm)\cite{galluzzi2019transport}. Despite this large separation, the conductance plateau remains well defined and reproducible. This length scale disfavors bulk-mediated nonlocal mechanisms in conventional superconductors, such as crossed Andreev reflection (CAR) or electron cotunneling (EC), whose amplitudes decay on the scale of $\xi$\cite{zeng2006local,russo2005experimental,cadden2006nonlocal,morten2006circuit}. Moreover, the bias-independent plateau observed over micrometer distances suggests that scattering and decoherence arising from this boundary-mediated transport process are strongly suppressed within the relevant low-energy window.

\section{\label{sec:Results5}Temperature evolution and correlation with magnetic order}
Temperature evolution provides a stringent test of whether the nonlocal plateau is resilient to ordinary decoherence mechanisms. For conventional AR spectra, increasing temperature rapidly induces thermal broadening and smearing of gap-related features, and the conductance anomalies continuously disappear as $T\rightarrow T_c$.\cite{daghero2011directional,deutscher2005andreev} By contrast, if the plateau originates from a protected boundary-mediated transport, its low-energy response is expected to exhibit markedly reduced thermal smearing over an extended temperature range, and may vanish at a temperature scale set by the additional order that stabilizes the boundary channel. 

In addition, \ch{FeTe_{0.55}Se_{0.45}} hosts three intertwined ingredients that can be relevant to the temperature dependence of our measurements: the nontrivial topological band structure, bulk superconductivity, and spontaneous magnetic order below $T_{Kerr}$. In the preceding sections, we established that the plateau requires a topologically nontrivial composition and a well-defined continuous edge, indicating the involvement of boundary electronic states. Now, we study the temperature dependence of the plateau to clarify which remaining ingredient controls its emergence. If the plateau were governed solely by superconductivity, its characteristic scale should track the superconducting gap and vanish at $T_c$; in contrast, if it is stabilized by magnetic order, its disappearance temperature and spectral evolution may instead correlate with $T_{Kerr}$. 

\begin{figure*} [t]
    \centering
    \includegraphics[width=1\linewidth]{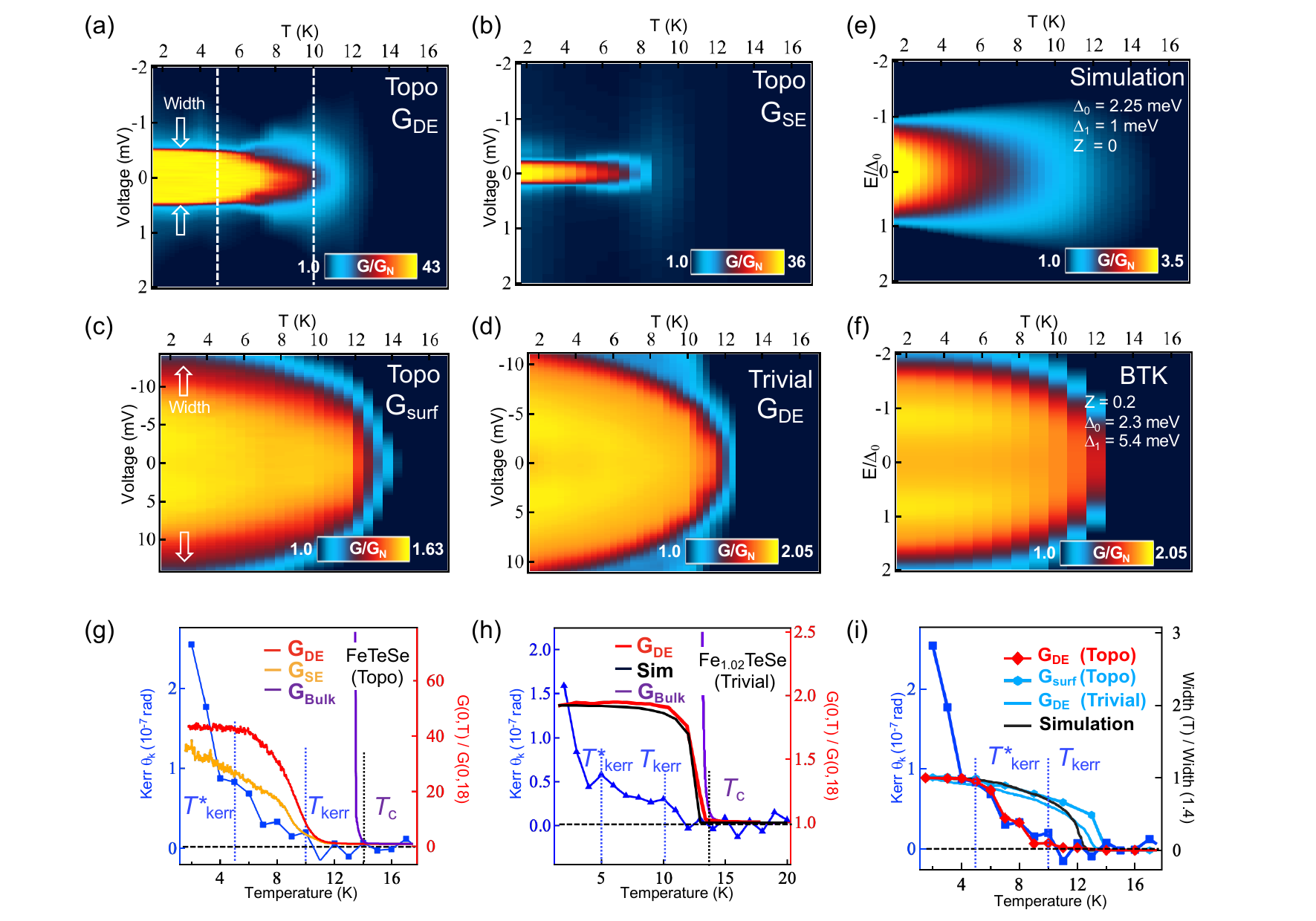}
    \caption{\label{fig:Tsweep} Temperature evolution of edge-related and surface differential conductance. (a) Temperature-dependent $G_{DE}$ measured on a \ch{FeTe_{0.55}Se_{0.45}} device, showing the evolution of the conductance plateau with temperature. (b) Temperature-dependent $G_{SE}$ measured on the same \ch{FeTe_{0.55}Se_{0.45}} device, exhibiting a ZBCP that evolves with temperature. (c) Temperature-dependent $G_{\mathrm{Surf}}$ measured on the same \ch{FeTe_{0.55}Se_{0.45}} device. (d) Temperature-dependent differential conductance $G_{DE}$ measured on a \ch{Fe_{1.02}Te_{0.55}Se_{0.45}} device. (e) Simulated temperature evolution of spectra in the perfect-Andreev-reflection limit for a \ch{FeTe_{0.55}Se_{0.45}} device. (f) Simulated temperature evolution of trivial Andreev-reflection spectra based on the Blonder–Tinkham–Klapwijk (BTK) model. (g) Comparison of the zero-bias conductance extracted from $G_{DE}$ and $G_{SE}$ with the polar Kerr rotation $\theta_{K}$ and the four-terminal bulk conductance $G_{\mathrm{Bulk}}$ measured on the \ch{FeTe_{0.55}Se_{0.45}} device. The onset of the Kerr signal occurs at a temperature distinct from the superconducting transition temperature $T_{c}$. (h) Same comparison as in (g), performed on the \ch{Fe_{1.02}Te_{0.55}Se_{0.45}} device. (i) Temperature dependence of the normalized spectral width extracted from different measurement configurations, compared with BTK simulations.}
\end{figure*}

Figure~\ref{fig:Tsweep}(a) shows the temperature-dependent differential conductance $G_{DE}(V,T)$ measured on a topological \ch{FeTe_{0.55}Se_{0.45}} device in the double-edge-lead configuration. The plateau evolution can be divided into three regimes (white dashed lines). For $T<5$~K, the plateau is nearly temperature independent, despite its sub-meV voltage window. This robustness is notable because conventional local spectra are expected to exhibit appreciable thermal smearing on such a small energy scale. Indeed, Fig.~\ref{fig:Tsweep}(e) presents the simulated temperature evolution for the perfect-Andreev case ($Z=0$) with bulk gap $\Delta_0 = 2.25$~meV, Dirac-surface-state gap $\Delta_1 = 1$~meV, and $T_c = 14$~K (see Supplemental Materials). Even in this transparent-interface limit, the local response is not protected against thermal broadening and quasiparticle scattering; consequently, the simulation predicts that a low-energy conductance enhancement is strongly reduced upon warming from $\sim 2$~K while the remaining gap-related features evolve continuously up to $T_c$. In contrast to this prediction, the experimental plateau remains essentially unchanged below 5 K. Then, upon entering the intermediate regime ($5$~K$<T<10$~K), the plateau width progressively narrows, and in the high-temperature regime it rapidly diminishes and vanishes near $T\approx 10$~K, well below $T_c=14.2$~K. On the same device, the ZBCP measured in the single-edge-lead configuration [$G_{SE}(V,T)$; Fig.~\ref{fig:Tsweep}(b)] also broadens and is suppressed with increasing temperature, disappearing near $10$~K.

In contrast, surface-contact measurements $G_{Surf}(V,T)$ on the same \ch{FeTe_{0.55}Se_{0.45}} flake [Fig.~\ref{fig:Tsweep}(c)] display conventional AR spectra. The AR features (within $\sim 10$~meV) are immediately thermally smeared and continuously vanish as $T \rightarrow T_c$, consistent with previous experiments\cite{park2010strong,tang2019quasi}. A similar temperature evolution is observed of $G_{DE}(V,T)$ in trivial \ch{Fe_{1.02}Te_{0.55}Se_{0.45}} devices [Fig.~\ref{fig:Tsweep}(d)], where no plateau is present. We note that the temperature evolution of these trivial cases can be well reconstructed within a two-gap BTK framework [Fig.~\ref{fig:Tsweep}(f), see Supplemental Materials], where the conductance features follow the superconducting gap scale and vanish continuously as $T \rightarrow T_c$. In addition, although trivial Andreev spectra begin to smear at relatively low temperatures, they persist up to $T_c$, unlike the plateau, which collapses already around 10 K, indicating that the plateau is not governed solely by bulk superconductivity. Thus, the plateau in topological devices exhibits two unconventional thermal characteristics: (i) its amplitude and width remain nearly unchanged below 5 K; (ii) it vanishes at 10 K, rather than $T_c$. 

Next, we turn to clarify the relationship between these two characteristics and the magnetism in this material. Specifically, in Fig.~\ref{fig:Tsweep}(g) we compare the normalized zero-bias conductance $G(0,T)$ extracted from $G_{DE}$ and $G_{SE}$ with the polar Kerr rotation $\theta_{K}$ and the four-terminal bulk conductance $G_{Bulk}$ measured on the same \ch{FeTe_{0.55}Se_{0.45}} device. $G_{Bulk}$ becomes infinite below $T_c \approx 14.2$ K, while $\theta_{K}$ onsets at $T_{Kerr} \approx 10$ K. Strikingly, both the plateau ($G_{DE}$) and the ZBCP ($G_{SE}$) disappear at $T_{Kerr}$. Furthermore, below $T_{Kerr}^*$ ($\sim$ 5 K), $G_{DE}(0,T)$ is nearly temperature-independent, mirroring the saturation behavior of the magnetization (see Supplemental Materials). The coincidence of these temperature scales indicates that the edge-associated conductance is additionally governed by magnetic order. As a control experiment, Fig.~\ref{fig:Tsweep}(h) shows the temperature evolution of $G(0,T)$ in trivial \ch{Fe_{1.02}Te_{0.55}Se_{0.45}} devices. There, the experimental $G(0,T)$ (red line) follows the BTK simulation (black line), vanishes at $T_c \approx 13$ K, and has no anomalies near $T_{Kerr}$ or $T_{Kerr}^*$. 

Finally, Fig.~\ref{fig:Tsweep}(i) compares the normalized spectral widths. In trivial configurations, the spectral width predicted by BTK simulations follows a BCS-like temperature evolution of the superconducting gap (see Supplemental Materials). In contrast, the plateau width closely tracks $\theta_{K}$: it remains constant below $T_{Kerr}^*$, shrinks for $T_{Kerr}^* < T < T_{Kerr}$, and collapses near $T_{Kerr}$. These observations establish a clear separation of energy scales: trivial Andreev features are governed by bulk superconductivity and vanish at $T_c$, whereas the conductance plateau is controlled by magnetic order and disappears at $T_{Kerr}$. The plateau, therefore, emerges only in the regime where superconductivity and spontaneous magnetization coexist.

\section*{Discussion}
Across multiple devices and measurement implementations, we identify a conductance plateau that appears only in topological \ch{FeTe_{0.55}Se_{0.45}} flakes when contacts address straight and continuous (sharp) edges. This behavior is reproducible across more than 10 topological devices with sharp-edge contacts (see Supplementary Material), and it persists across substantial variations in experimental setups, including in a separate laboratory, different-metal contact (e.g., Al), and T-SPL based devices with nanometer-scale electrodes. Such reproducibility already argues against an origin tied to a specific contact metal, fabrication protocol, or measurement environment.

Additional checks further argue against an origin from measurement artifacts, contact non-idealities, or edge-induced inhomogeneity. First, the plateau is invariant across measurement schemes: it is reproduced under both DC and AC excitation, and across two- and three-terminal configurations using various contacts on the same edge [Fig.~\ref{fig:CircuitFTS}(b,c)]. Second, a contact-size analysis based on the measured $R_{N}$ places our contacts in the ballistic or near-ballistic regime, with effective sizes comparable to the flake thickness, disfavoring non-spectroscopic zero-bias anomalies that arise in strongly non-ballistic contacts\cite{sheet2004role,gifford2016zero}. Third, cross-sectional ADF-STEM reveals a residual oxidized layer that suppresses direct top-surface coupling beneath the metal while enabling robust side-surface coupling, consistent with an edge-contact measurement geometry. Finally, EDX elemental maps show spatially uniform Fe/Te/Se distributions along the edges, ruling out compositional phase segregation at rough or sharp boundaries. Taken together, these checks indicate that the plateau reflects an intrinsic transport response tied to edge coupling.

A primary question is whether the edge plateau or ZBCP can be explained by known non-topological transport mechanisms in conventional superconductors. One possibility is that the observed zero-bias features arise from conventional edge-localized Andreev bound states (ABS), which commonly produce a ZBCP in local tunneling spectra. In addition, spatially separated contacts can exhibit a finite nonlocal signal through crossed Andreev reflection (CAR) and elastic electron cotunneling (EC), where the source and drain contacts are coupled via bulk Cooper pairs\cite{russo2005experimental,cadden2006nonlocal,morten2006circuit}. A defining feature of such bulk-mediated processes is that their amplitudes decay on the scale of the superconducting coherence length.

However, several aspects of our data are distinguishable from these conventional pictures: (i) The plateau/ZBCP response is strongly drain dependent, namely, while keeping the source junction unchanged, moving the drain from a double-edge to a single-edge configuration converts the spectrum from a conductance plateau to a ZBCP. This behavior indicates a nonlocal edge response, inconsistent with a purely local ABS determined only by the injection contact. (ii) The nonlocal response extends over micrometer-scale source--drain separations, far exceeding the bulk coherence length of Fe(Te,Se) ($\sim 6$ nm)\cite{galluzzi2019transport}, thereby excluding conventional bulk-mediated CAR/EC as the dominant origin. (iii) the phenomenology exhibits a combined dependence on sample topological phase and edge geometry. In topologically trivial \ch{FeTe_{0.40}Se_{0.60}} and \ch{Fe_{1.02}Te_{0.55}Se_{0.45}}, only conventional double-peak Andreev spectra are observed irrespective of whether the contacted edge is sharp or rough, whereas in topologically nontrivial \ch{FeTe_{0.55}Se_{0.45}}, sharp continuous edges produce the plateau and rough or stepped edges yield a ZBCP. (iv) The suppression temperature of both the plateau and ZBCP is strongly correlated with the TRSB temperature scale rather than the bulk superconducting transition temperature $T_c$, which cannot be explained under the conventional Andreev scenario. (v) The plateau maintains strongly suppressed thermal broadening below $T_{Kerr}^*$, whereas neither local trivial ABS nor bulk CAR/EC is expected to be topologically protected against thermal broadening. These observations suggest that the edge plateau reflects a boundary-mediated superconducting transport mechanism beyond conventional local ABS or bulk-state CAR/EC.

Instead, our results point to boundary-mediated nonlocal and protected conduction emerging from the interplay of topology, superconductivity, and spontaneous TRSB in \ch{FeTe_{0.55}Se_{0.45}}. To account for the full set of constraints simultaneously, a natural phenomenological framework is that the edge hosts topological boundary excitations that couple to normal-metal leads in a fundamentally different way from ordinary quasiparticles. Although such boundary excitations can be charge-neutral on average, multi-terminal charge transport can arise when they are contacted by normal-metal electrodes: charge is injected and extracted through Andreev-type conversion processes at the terminals, while an edge-mediated phase-coherent propagation pathway connects spatially separated contacts (see Supplemental Materials).\cite{zhang2017quantum,law2009majorana,he2014correlated} Within this view, a plateau-like conductance can arise when the effective edge-mediated transmission is only weakly energy dependent over a finite low-energy window. The same framework naturally accounts for the strong drain-location sensitivity and micrometer-scale range, since the measured signal depends on whether source and drain are connected by an uninterrupted edge segment.\cite{he2014correlated,zhang2017quantum,ikegaya2019anomalous,casas2024long} 

In this scenario, the ZBCP observed in rough-edge contacts may reflect crystal steps or edge discontinuities that interrupt the continuous side-surface boundary channel. However, the microscopic mechanism by which such interruptions convert an extended edge-mediated plateau into a localized ZBCP remains unresolved. In addition, the absolute magnitude of the plateau also remains beyond a simple quantized single-channel description. In particular, the measured conductance is substantially larger than the expectation for a single transmitted boundary channel, suggesting that multi-channel transport and three-dimensional/multiband effects are likely essential. Moreover, whether bulk-state Andreev processes or conversion between boundary modes and bulk superconducting states also contribute to the measured conductance is presently unclear. A quantitative description of the plateau magnitude therefore requires theoretical modeling beyond the existing minimal scenarios. In addition, experimental studies of a wide range of Fe(Te,Se) flake thicknesses and contact geometries will be crucial in evaluating specific scenarios.

The remaining question is which topological phase with boundary channels is responsible for these observations? For a 3D superconductor with a nontrivial electronic structure, TRSB in \ch{FeTe_{0.55}Se_{0.45}} could enable at least two broad scenarios for chiral boundary modes. First, the TRSB order could split bulk topological bands into a Weyl regime under suitable exchange energy and chemical potential conditions, yielding a Weyl superconductor with chiral boundary modes analogous to Fermi-arc-derived channels\cite{sau2012topologically,meng2012weyl}. Second, bulk symmetry breaking may generate an effectively layered (stacked) chiral topological superconducting phase, analogous to 3D chiral $p$-wave superconductivity\cite{hughes2014majorana,wu2021topological,hu2024dislocation}. Our experiments impose a key constraint on any such scenario: both the plateau and the edge-contact ZBCP vanish at $T_{Kerr}$, well below $T_c$ [Fig.~\ref{fig:Tsweep}(g)], and the plateau width tracks $\theta_{K}$ across temperature [Fig.~\ref{fig:Tsweep}(i)]. This correlation indicates that the low-energy transport scale associated with the plateau is not set solely by $\Delta_{SC}$, but is instead governed by an energy scale tied to TRSB order.

Our work establishes experimental criteria that collectively support an edge-state interpretation. Nonetheless, further theoretical and experimental efforts are needed to identify the specific superconducting state (e.g., Weyl versus stacked chiral) responsible for the boundary response in this single-material platform. On the theory side, while the observed nonlocal phenomenology is broadly consistent with models of transport mediated by topological superconducting edge modes\cite{he2014correlated,zhang2017quantum,ikegaya2019anomalous}, quantitative comparison including the bias independent conductance, its temperature, value, and drain dependence will likely require extensions to multi-channel or fully 3D geometries, and to the multiband superconductivity of Fe(Te,Se)\cite{mascot2022topological}. Furthermore, the role of the bulk in transport must be clarified. Experimentally, phase-sensitive interferometry (edge Josephson interferometers or SQUID geometries engineered along a single crystallographic edge) could directly probe the phase structure of the edge conduction channel. Gate or etch-defined edge constrictions and quantum point contacts may enable controlled mode filtering and tests of backscattering suppression. Complementary thermal-transport measurements along device edges would provide an orthogonal probe to charge transport, and correlating edge transport with controlled TRSB-domain manipulation and spatially resolved Kerr imaging could clarify how magnetic order organizes the boundary channel. More broadly, elucidating the role of correlations in generating the intertwined magnetic, superconducting, and topological orders in \ch{FeTe_{0.55}Se_{0.45}} remains an important open direction\cite{Kim.OrbitalSelectiveFTS.2024}.

\begin{acknowledgments}
We thank Fan Zhang, Ilija Zeljkovic, Ziqiang Wang, and Jigang Wang for helpful discussions. Disclaimer by NIST: Certain commercial equipment, instruments, software, or materials are identified in this paper in order to specify the experimental procedure adequately; Such identifications are not intended to imply recommendation or endorsement by NIST, nor are they intended to imply that the materials or equipment identified are necessarily the best available for the purpose.
\noindent\textbf{Funding:} 
The National Science Foundation, Award No., supported the work of W.L., M.Gray, and J.V. MRI-2117711, DMR-2310895. The fabrication efforts of M.Geiwitz, G.N. and V.L. was supported by the U.S. Department of Energy (DOE), Office of Science, Basic Energy Sciences (BES) under Award \#SC0018675. The AFSOR, Grant FA9550-20-1-0282, supported the work of K.S.B. Transport measurements at Boston College were enabled by equipment provided through AFOSR DURIP award FA9550-20-1-0246. The Gordon and Betty Moore Foundation EPiQS Initiative, Grant\#GBMF10276 and and NSF award DMR-2419425 supported the work at UC Irvine. The work at BNL was supported by the US Department of Energy, oﬃce of Basic Energy Sciences, contract no. DOE-sc0012704. Q.M. and V.B. acknowledge support from the NSF CAREER award DMR-2143426 as well as the Alfred P. Sloan Foundation. J.C. and J.H. acknowledge support from the Gordon and Betty Moore Foundation EPiQS Initiative, Grant\#GBMF9062.01. H.Z. acknowledges support from NIST Cooperative Agreement\#70NANB22H101. E.R. acknowledges support from the US Department of Energy, Office of Basic Energy Sciences, via Award DE-SC0022245. The NSF\/DMR-2003405, NSF\/DMR-1644779, and the State of Florida supported the work at NHMFL and FSU. K.W. and T.T. acknowledge support from the JSPS KAKENHI (Grant Numbers 21H05233 and 23H02052) , the CREST (JPMJCR24A5), JST and World Premier International Research Center Initiative (WPI), MEXT, Japan. Q.T. and X.L. acknowledge the support by the National Science Foundation (NSF) under Grant No. 1945364 and US Department of Energy (DOE), Office of Science, Basic Energy Sciences (BES) under Award DE-SC0021064. E.R., R.Z. and K.S.B performed part of his work while at the Kavli Institute of Theoretical Physics (KITP) that is supported by the National Science Foundation Grant No. NSF PHY-1748958.
\noindent \textbf{Author contributions:} K.S.B. conceived and supervised the project; W.L., G.N. and K.S.B. designed and conducted the differential conductance experiments, with the help of M.G., E.A., and K.F.; Exfoliation and AFM characterization was performed by W.L., G.N., M.G., V.L., J.V., V.B., and Q.M.; The h-BN/FTS devices were fabricated by G.N., and W.L.; C.F. and J.X. conducted the Kerr experiments and analysis; W.L. and G.N. analyzed the differential conductance data with help from W.K.P.; The theoretical discussion is contributed by R.Z. and E.R.; E.R., R.Z., and X.G. performed the calculations; G.G. provided the FeTeSe materials; T.T. and K.W. provided the h-BN materials; J.H. and J.C. performed the STEM measurement and EDS mapping; Q.T. and X.L. performed the EDS measurement; H.Z. and A.D. performed the ADF-STEM measurement; W.L. and G.N. drafted the manuscript with the help of J.X. and K.S.B.; All authors contributed to the discussion of the manuscript. 
\end{acknowledgments}

\noindent \textbf{Data and materials availability:} All data are available in the manuscript or the Supplemental Materials.

\section*{Competing interests} 
The authors declare no competing interest. 

\noindent\textbf{Supplemental Materials}

\nocite{*}

\bibliography{Reference.bib}

\end{document}